\documentclass{article}

\usepackage{arxiv}

\usepackage[utf8]{inputenc} 
\usepackage[T1]{fontenc}    
\usepackage{hyperref}       
\usepackage{url}            
\usepackage{booktabs}       
\usepackage{amsfonts}       
\usepackage{nicefrac}       
\usepackage{microtype}      
\usepackage{lipsum}		
\usepackage{graphicx}
\usepackage{natbib}
\usepackage{doi}
\usepackage{xspace}
\newcommand{\hn}[0]{\textsc{hn}\xspace}
\newcommand{\hns}{\textsc{hn}s\xspace}

\title{MeVer at CheckThat! 2026: Cluster-Aware Hard-Negative Mining for Multilingual Scientific Source Retrieval}

\author{
Juli Bakagianni\\
Information Technologies Institute,\\
Centre for Research and Technology Hellas (CERTH)\\
Thessaloniki, Greece\\
\texttt{julibak@iti.gr}
\And
Symeon Papadopoulos\\
Information Technologies Institute, \\
Centre for Research and Technology Hellas (CERTH)\\
Thessaloniki, Greece\\
\texttt{papadop@iti.gr}
}

\date{}

\renewcommand{\headeright}{CheckThat! 2026}
\renewcommand{\undertitle}{CLEF 2026 Working Notes}
\renewcommand{\shorttitle}{MeVer at CheckThat! 2026}

\begin{document}
\maketitle

\begin{abstract}
Identifying the scientific source behind a social media claim requires matching short, informal, and often multilingual claims against large collections of scientific publications, where semantically related papers may act as challenging distractors or false negatives during training. We present our submission to CheckThat! 2026 Task 1 on multilingual scientific source retrieval, focusing on how hard-negative mining should be adapted to multi-stage retrieval pipelines for scientific source retrieval. We propose cluster-aware hard-negative mining strategies that exploit the semantic structure of retrieved candidate pools in order to construct more informative training negatives for dense retrieval and reranking. Our experiments show that different hard-negative structures induce different retrieval behaviors. Localized cluster negatives tend to favor precision-oriented retrieval, whereas broader non-gold semantic negatives provide stronger candidate coverage and more consistent reranking performance across languages. We further study multiple LLM-based evidence selection formulations, including direct classification, pairwise comparison, and listwise reranking prompts, and find that constrained classification prompts provide the most reliable final document selection. The final system combines a dense retriever, a multilingual cross-encoder reranker, and a selective LLM-based disagreement resolver, ranking 6th among 37 submissions in the shared task evaluation. Overall, our results suggest that hard-negative mining should be treated as a stage-aware design problem rather than as a single retrieval optimization strategy.
\end{abstract}

\begin{keywords}
information retrieval; social media retrieval; hard-negative mining; scientific source retrieval; cross-encoder reranking; large language models
\end{keywords}

\maketitle

\section{Introduction}

Scientific claims are frequently discussed on social media through informal references to studies, reports, or ``research findings'' without explicit citations or links to the underlying publication. Automatically identifying the referenced scientific paper is therefore an important problem for scientific fact-checking, evidence tracing, and misinformation analysis \cite{clef-checkthat:2026-lncs, clef-checkthat:2026:task1, hafid2026scientific}. However, this retrieval setting is particularly challenging because short and loosely phrased social media claims must be matched against large collections of scientific articles written in formal academic language. The problem becomes even more difficult in multilingual settings, where claims may appear in different languages while the scientific collection remains primarily English. Moreover, since the dataset provides only a single annotated source paper per claim, semantically related but unlabeled papers may act as false negatives during retrieval and reranking training.

CheckThat! Task 1, organized as part of the CLEF CheckThat! Lab \cite{clef-checkthat:2026-lncs}, formulates this problem as a multilingual scientific source retrieval task. Given a social media claim in English, German, or French, the goal is to retrieve the referenced scientific paper from a shared collection of scientific publications \cite{clef-checkthat:2026:task1}. The task combines known-item retrieval with semantic evidence matching: the correct paper may differ from competing candidates only in subtle methodological or semantic details, requiring systems to balance broad candidate coverage with fine-grained semantic discrimination.

Modern retrieval systems for such tasks are commonly organized as multi-stage pipelines. First-stage dense retrievers retrieve candidate documents using learned embedding similarity representations \cite{karpukhin2020dense, xiong2020approximate}. Cross-encoder rerankers then refine the ranking through full query–document interaction modeling \cite{nogueira2019passage, qu2021rocketqa}. More recently, large language models (LLMs) have also been explored as reasoning-based rerankers or judges capable of comparing a small set of candidate documents and selecting the best supporting evidence \cite{sun2023chatgpt, qin2024large}. These stages address complementary objectives: retrieval emphasizes candidate-set coverage, reranking emphasizes local semantic discrimination, and LLM-based judgment targets difficult final selection cases where multiple candidates remain plausible.

In this work, we focus on how hard negative (\hn) mining during model training should be adapted to such a multi-stage retrieval pipeline. Existing \hn approaches for contrastive retriever and reranker training typically select negatives based on retrieval rank or nearest-neighbor similarity \cite{xiong2020approximate, zhan2021optimizing}. However, retrieved candidate pools often exhibit internal semantic structure: some documents form highly localized neighborhoods around the gold paper, while others act as broader topical distractors. We therefore investigate whether clustering retrieved candidates can provide a more principled mechanism for selecting informative training negatives for both retriever and reranker fine-tuning.

Our central hypothesis is that different \hn structures induce different retrieval behaviors across multi-stage pipelines. Broader semantic negatives may improve downstream candidate coverage, whereas highly localized negatives may sharpen fine-grained semantic discrimination between closely related scientific papers. We therefore study how different cluster-aware \hn strategies affect both retrieval and reranking stages. To study this question, we introduce cluster-aware \hn mining strategies that sample negatives from the gold-document cluster, neighboring clusters, or non-gold semantic regions of the retrieved space.

Beyond retrieval and reranking, we also investigate how LLMs should operate as selective final-stage judges. Rather than applying the LLM universally, we use it only on disagreement cases between the retriever and reranker. We further study multiple prompting and decision formulations for LLM-based evidence selection, including direct classification, pairwise comparison, permutation-style ranking inspired by RankGPT \cite{sun2023chatgpt}, shuffled candidate ordering, and committee-style voting. Our experiments show that simple constrained classification prompts are more reliable than more elaborate ranking-style formulations in this retrieval setting.

Although multilingual handling is not the primary focus of this work, our experiments also reveal an interesting interaction between translation strategy and judge model family. Translation into English consistently improves first-stage retrieval and reranking for non-English queries, likely because the scientific collection itself is English. However, at the LLM judgment stage, GPT-family models benefit from the original-language claim formulation, whereas Llama-based judges perform better with translated claims. This suggests that optimal query representation depends not only on the pipeline stage but also on the downstream judge model and the candidate distribution produced by earlier retrieval stages.

The paper is organized around the following research questions:
\begin{itemize}
    \item Can cluster-aware \hn mining improve retrieval candidate coverage and reranking effectiveness in multilingual scientific source retrieval?
    \item Do retrieval and reranking stages benefit from different forms of semantic \hn?
    \item How should LLMs be used as final-stage judges in retrieve-then-rerank pipelines, and which prompting formulations are most effective for evidence selection?
\end{itemize}

Our final system combines a dense retriever and a cross-encoder reranker trained with cluster-aware \hn and a selective GPT-5.5 disagreement resolver. Across experiments, we find that non-gold-cluster negatives provide the strongest retrieval backbone for reranking-oriented pipelines, while localized cluster negatives are more effective for sharpening retrieval-only precision. We further show that selective LLM intervention is most effective when restricted to disagreement cases and formulated as constrained classification rather than free-form ranking. Together, these results suggest that \hn mining should be treated as a stage-aware design problem rather than a single retrieval optimization strategy.

\section{Related Work}

Dense retrieval has become a standard approach for semantic document retrieval using contrastive representation learning \cite{karpukhin2020dense, xiong2020approximate}. A central component of dense retriever training is \hn mining, where models are trained to distinguish relevant documents from semantically similar distractors. Prior work has explored Approximate Nearest Neighbor Negative Contrastive Learning (ANCE) \cite{xiong2020approximate}, BM25-based negatives \cite{karpukhin2020dense}, denoised \hn \cite{zhan2021optimizing}, and curriculum-style training strategies. However, most existing approaches define negative difficulty primarily through retrieval rank or embedding similarity, without explicitly modeling the semantic structure of the retrieved candidate space.

Retrieve-then-rerank pipelines further improve retrieval quality by combining dense retrievers with cross-encoder rerankers \cite{nogueira2019passage, qu2021rocketqa}. More recently, LLMs have also been studied for ranking or reranking tasks \cite{sun2023chatgpt, qin2024large}, including permutation-based ranking and pairwise prompting strategies. Existing work has shown that LLMs can act as effective rerankers over small candidate sets, although their reliability and prompting sensitivity remain active research questions.

\section{Dataset}

We use the official \textit{CheckThat! 2026 Task 1} dataset for multilingual scientific source retrieval.\footnote{\url{https://huggingface.co/datasets/sschellhammer/CT26_Task1_SourceRetrievalForScientificWebClaims}} The dataset consists of social media claims from X in English, German, and French paired with a shared scientific document collection containing 10,000 English-language scientific papers. Each query contains a scientific claim together with an implicit reference to a scientific paper, and the task is to retrieve the referenced paper from the candidate collection \cite{clef-checkthat:2026:task1}. Each query is associated with a single annotated gold scientific paper. Each paper includes title, abstract, venue, and author metadata. Table~\ref{tab:dataset_examples} presents examples of social media posts with implicit references to scientific articles, while Table~\ref{tab:publication_examples} shows example papers from the scientific collection.

\begin{table*}[ht!]
\centering
\caption{Example social media claims with implicit references to scientific publications from the CheckThat! 2026 Task 1 dataset.}
\label{tab:dataset_examples}
\small
\resizebox{\textwidth}{!}{
\begin{tabular}{p{0.06\textwidth} p{0.78\textwidth} p{0.08\textwidth}}
\toprule
Lang & Tweet Text & Pub ID \\

\midrule

EN &
Published in the journal \textit{Antiviral Research}, the study from Monash University showed that a single dose of Ivermectin could stop the coronavirus growing in cell culture -- effectively eradicating all genetic material of the virus within two days.
& 234567 \\

\midrule

DE &
Die in der Fachzeitschrift \textit{Antiviral Research} veröffentlichte Studie der Monash University zeigte, dass eine einzige Dosis Ivermectin das Wachstum des Coronavirus in Zellkulturen stoppen kann und das gesamte genetische Material des Virus innerhalb von zwei Tagen effektiv zerstört.
& 234567 \\

\midrule

FR &
L'outil génétique CRISPR-Cas9 a été utilisé pour permettre à la provitamine D3 de s'accumuler dans les feuilles et fruits de la tomate. Une fois exposée aux UV, elle s'est bien convertie en vitamine D (dont 1 milliard de personnes est carencé dans le monde).
& 345678 \\

\bottomrule
\end{tabular}
}
\end{table*}

\begin{table*}[ht!]
\centering
\caption{Example scientific publications from the shared document collection of CheckThat! 2026 Task 1.}
\label{tab:publication_examples}
\small
\resizebox{\textwidth}{!}{
\begin{tabular}{p{0.08\textwidth} p{0.28\textwidth} p{0.28\textwidth} p{0.30\textwidth}}
\toprule
Pub ID & Study Title & Study Venue + Authors & Study Abstract \\
\midrule

123456 &
\textit{Effectiveness of Covid-19 Vaccines against the B.1.617.2 (Delta) Variant}
&
[Venue]: \textit{New England Journal of Medicine} [Authors]:
Jamie Lopez Bernal, Nick Andrews, Charlotte Gower, Eileen Gallagher, Ruth Simmons, Simon Thelwall, Julia Stowe, Elise Tessier, \ldots
&
BACKGROUND: The B.1.617.2 (delta) variant of the severe acute respiratory syndrome coronavirus 2 \ldots \\

\midrule

234567 &
\textit{The FDA-approved drug ivermectin inhibits the replication of SARS-CoV-2 in vitro}
&
[Venue]: \textit{Antiviral Research} [Authors]:
Caly, Leon; Druce, Julian D.; Catton, Mike G.; Jans, David A.; Wagstaff, Kylie M.
&
Although several clinical trials are now underway to test possible therapies, the worldwide response to the COVID-19 outbreak has been largely limited to monitoring/containment \ldots \\

\bottomrule
\end{tabular}
}
\end{table*}

In the collection, paper titles are short (13 words on average), whereas abstracts are much longer (232 words on average). Table~\ref{tab:dataset_stats} summarizes the query distribution across train, development, and test splits. The dataset is heavily English-dominant, while German and French constitute substantially smaller portions of the benchmark.

\begin{table}[ht!]
\centering
\small
\caption{Query distribution across dataset splits.}
\label{tab:dataset_stats}
\begin{tabular}{lccc}
\toprule
Split & English & German & French \\
\midrule
Train & 14,977 & 1,460 & 2,807 \\
Dev & 3,905 & 386 & 702 \\
Test & 6,076 & 876 & 1,220 \\
\bottomrule
\end{tabular}
\end{table}

\section{Methodology}

Our system follows a three-stage pipeline. First, a dense retriever selects a candidate pool of potentially relevant scientific papers. Second, a cross-encoder reranker refines this candidate set through finer query--document interactions. Finally, an LLM judge is applied selectively to disagreement cases where the retriever and reranker produce different top predictions.

The main methodological focus of this work is cluster-aware \hn mining during retriever and reranker training. We study whether the semantic structure of retrieved candidate pools can be exploited through clustering to construct more informative training negatives for different pipeline stages. In addition, we investigate multilingual query handling and multiple LLM-based evidence selection formulations for final stage document selection.

\subsection{Multilingual Query Handling}

The scientific document collection used in the task is entirely English, while claims are provided in English, German, and French. We examine multilingual handling at two stages of the pipeline. The first is the retrieval stage, where the query is matched against English scientific documents. Translation may help by aligning the claim language with the document collection and simplifying semantic matching. At the same time, translation may also remove linguistic nuances or domain-specific phrasing that could remain useful for later evidence selection.

The second stage is the LLM-based judgment component, where the model receives the claim together with a small set of candidate papers and must select the strongest supporting source. Here, we study whether the judge should operate on the original language claim or on its English translation. We therefore treat multilingual query handling as a methodological design choice rather than a fixed preprocessing step.

\subsection{Dense Retrieval Stage}

The first stage of the pipeline is a dense retriever trained to retrieve a candidate pool of relevant scientific papers for each claim. Retriever training follows the standard contrastive learning setting, where the model is optimized to assign higher similarity to the gold paper than to negative documents. Since retrieval effectiveness depends strongly on the quality and difficulty of the training negatives, we investigate multiple \hn mining strategies.

All training settings use standard in-batch negatives, where negatives are formed from the positive documents of other examples within the same training batch. In-batch negatives provide a large number of inexpensive and diverse easy negatives that stabilize contrastive retriever training \cite{karpukhin2020dense, xiong2020approximate}. 

As a stronger \hn baseline, we study ANCE \hn \cite{xiong2020approximate}, where the current retriever periodically retrieves top-ranked candidate documents that are then used as \hns for subsequent contrastive training. Our cluster-aware approaches build directly on this setting: rather than sampling \hn only according to retrieval similarity, we additionally exploit the semantic structure of the retrieved candidate pool through clustering.

Starting from the retrieved candidate pool of the current retriever, we cluster candidate document embeddings and construct different types of training negatives. Let the \textit{gold cluster} denote the cluster containing the gold document. We study three cluster-aware strategies:

\begin{itemize}
    \item \textbf{Gold-cluster negatives}: negatives are sampled from the same cluster as the gold document, forcing the retriever to distinguish between highly similar local distractors.
    
    \item \textbf{Nearest-cluster negatives}: negatives are sampled from the non-gold cluster whose centroid is nearest to the gold cluster centroid, targeting semantically neighboring but distinct documents.
    
    \item \textbf{Non-gold-cluster negatives}: negatives are sampled from retrieved documents outside the gold cluster, providing broader semantic diversity while remaining retrieval-relevant.
\end{itemize}

The motivation behind these variants is that different retrieval objectives may benefit from different forms of semantic supervision. Gold-cluster negatives emphasize fine-grained discrimination between closely related papers, whereas non-gold-cluster negatives encourage broader candidate coverage by exposing the retriever to more diverse semantic distractors. Since the downstream reranker can only operate on retrieved candidates, candidate coverage remains particularly important in our retrieve--then--rerank pipeline.

\subsection{Reranking Stage}

The second stage of the pipeline is a cross-encoder reranker that reorders the candidate documents returned by the dense retriever. Unlike dense retrieval, where queries and documents are compared through independent vector representations, the reranker jointly processes the query and candidate document, enabling finer-grained semantic matching between claim wording and scientific abstracts.

Reranker training operates under a different setting from retrieval training. While the retriever must separate the gold paper from the full document collection, the reranker only observes the restricted candidate pool produced by the retriever. 
Starting from the retrieved candidate pool, we apply the same cluster-aware \hn framework used in retrieval training. We compare negatives sampled from the gold cluster, the nearest non-gold cluster, and non-gold semantic regions of the retrieved space. In this setting, gold-cluster negatives emphasize fine-grained discrimination between highly similar scientific papers, whereas broader non-gold negatives expose the reranker to more diverse retrieval distractors.

\subsection{LLM-as-Judge Stage}

The final stage of the pipeline uses an LLM as a selective evidence selection judge. Rather than applying the LLM over all reranked candidates, we restrict its use to disagreement cases where the dense retriever and reranker produce different top-ranked predictions. This setting focuses the LLM on difficult cases where multiple plausible candidate papers survive the earlier pipeline stages.

The judge operates over a small candidate set constructed from the top reranked documents together with the retriever top-ranked document whenever it differs from the reranker prediction. In this way, the LLM acts as a targeted resolver between competing retrieval and reranking hypotheses rather than as a full reranker over the entire candidate pool.

Our main focus at this stage is the formulation of the evidence selection task itself. We investigate multiple prompting strategies:

\begin{itemize}
    \item \textbf{Direct classification}: the LLM directly selects the best supporting document from the candidate set.
    
    \item \textbf{Pairwise comparison}: the reranker top prediction is treated as a baseline candidate, and the LLM evaluates whether another candidate provides stronger evidence.
    
    \item \textbf{Listwise reranking}: the LLM produces a complete ordering over the candidate documents using permutation-style ranking prompts \cite{sun2023chatgpt}.
\end{itemize}

These formulations allow us to study whether LLMs operate more reliably as constrained classifiers, pairwise evaluators, or free-form ranking agents in scientific-source retrieval.

\section{Experimental Setup}

Our retrieval experiments use \texttt{BAAI/bge-large-en-v1.5}\footnote{\url{https://huggingface.co/BAAI/bge-large-en-v1.5}} as the main dense retriever. Preliminary retrieval experiments additionally included \texttt{multilingual-e5-large}\footnote{\url{https://huggingface.co/intfloat/multilingual-e5-large}} and \texttt{e5-large-v2}\footnote{\url{https://huggingface.co/intfloat/e5-large-v2}} (see Appendix~\ref{app:retr-models}). For reranking, we evaluated several cross-encoder families, including SciBERT and \texttt{BAAI/bge-reranker-base}, while the strongest reranking results were obtained with \texttt{jinaai/jina-reranker-v2-base-multilingual} (see Appendix~\ref{app:crossenc}).\footnote{\url{https://huggingface.co/jinaai/jina-reranker-v2-base-multilingual}} The LLM-based judgment stage was evaluated using \texttt{Llama 3 70B}, \texttt{gpt-4.1-mini}, and \texttt{gpt-5.5} (see Appendix~\ref{app:llm-comp}).

For clustering, we use the K-Means clustering algorithm. The number of clusters is selected by maximizing the macro silhouette score over a small search range ($k \in [3,6]$) \cite{pavlopoulos2024revisiting}. For retrieval training, we use a batch size of 16, yielding 15 in-batch negatives per example. In all \hn settings, a single additional \hn is sampled per query.
For reranking experiments, we use a top-200 retriever pool for \hn mining, exclude the gold document and the highest-ranked retrieved documents from negative sampling, and sample 10 \hn per query. Hyperparameters, including learning rate, training epochs, and \hn counts, are selected on the development set.

The official shared-task metric is MRR@5. Since our system follows a retrieve--then--rerank pipeline, we additionally track recall-oriented metrics during retrieval evaluation, particularly Recall@20, to ensure that the gold document remains available to downstream reranking and LLM-based selection stages.

\section{Results}

We first examine dense retrieval and reranking experiments with cluster-aware \hn mining, and finally LLM-based evidence selection ablations.

\subsection{Dense Retrieval Results}

Table~\ref{tab:retrieval_tuned} compares the negative mining strategies for dense retrieval. We evaluate in-batch negatives, ANCE \hn, and the proposed cluster-aware variants using Recall@20 as the primary metric, and then Recall@10 and MRR@5. Detailed multilingual query handling experiments are provided in Appendix~\ref{app:multi-handling}.

\begin{table*}[ht!]
\centering
\caption{Dense retrieval results for different \hn mining strategies on the development set. Metrics include Recall@20, Recall@10, and MRR@5 across languages and macro averages.}
\label{tab:retrieval_tuned}
\small
\resizebox{\textwidth}{!}{
\begin{tabular}{lcccccccccccc}
\toprule
& \multicolumn{3}{c}{EN} & \multicolumn{3}{c}{DE} & \multicolumn{3}{c}{FR} & \multicolumn{3}{c}{Macro} \\
\cmidrule(lr){2-4} \cmidrule(lr){5-7} \cmidrule(lr){8-10} \cmidrule(lr){11-13}
Family & R@20 & R@10 & MRR@5 & R@20 & R@10 & MRR@5 & R@20 & R@10 & MRR@5 & R@20 & R@10 & MRR@5 \\
\midrule
In-batch negatives only & \textbf{0.842} & 0.794 & \textbf{0.644} & 0.780 & 0.707 & 0.559 & 0.869 & 0.832 & 0.670 & 0.830 & 0.778 & 0.624 \\
ANCE & 0.838 & 0.792 & 0.635 & \textbf{0.824} & 0.764 & 0.589 & 0.877 & 0.846 & \textbf{0.715} & 0.847 & 0.801 & 0.646 \\
Gold-cluster negatives & 0.834 & 0.784 & 0.629 & 0.808 & 0.756 & \textbf{0.601} & 0.882 & 0.848 & 0.709 & 0.841 & 0.796 & 0.646 \\
Nearest-cluster negatives & 0.840 & 0.787 & 0.632 & 0.816 & 0.759 & 0.593 & 0.882 & 0.858 & 0.702 & 0.846 & 0.801 & 0.642 \\
Non-gold-cluster negatives & 0.839 & \textbf{0.800} & \textbf{0.644} & 0.819 & \textbf{0.769} & 0.590 & \textbf{0.886} & \textbf{0.865} & 0.707 & \textbf{0.848} & \textbf{0.811} & \textbf{0.647} \\
\bottomrule
\end{tabular}
}
\end{table*}

Overall, the differences between \hn strategies remain marginal to modest. One possible explanation is that strong in-batch supervision already provides substantial negative diversity, so the additional mined negatives affect only part of the training signal. However, this is not only a consequence of using a single explicit \hn: the scaling experiments in Section~\ref{sec:scaling} show that increasing the number of explicit \hn still does not substantially improve Recall@20. 

Despite these small margins, several consistent patterns emerge. First, standard in-batch negatives already provide a strong retrieval baseline, achieving the best English Recall@20. Since English also constitutes the largest portion of the training data, this suggests that large numbers of diverse in-batch negatives may already provide sufficiently strong supervision in higher-resource settings.

Second, the impact of \hn structure differs across the translated German and French subsets. While ANCE and cluster-aware approaches consistently improve over the in-batch baseline for these subsets, localized cluster negatives tend to produce stronger precision-oriented retrieval. In particular, gold-cluster negatives achieve the highest German MRR@5 and outperform the nearest and non-gold cluster variants on both German and French MRR@5. This suggests that localized semantic distractors may improve fine-grained ranking precision when retrieval supervision is weaker or noisier.

Finally, non-gold-cluster negatives achieve the strongest macro Recall@20 and Recall@10 scores, indicating that broader semantic negatives improve retrieval coverage across languages. Since downstream reranking and LLM-based selection can only operate on retrieved candidates, we therefore use the non-gold-cluster strategy as the retrieval backbone for subsequent reranking experiments.

\subsubsection{Scaling}\label{sec:scaling}

Table~\ref{tab:retrieval_scaling} examines the interaction between in-batch negatives and explicit \hn for the tuned non-gold-cluster retriever. We vary both the effective number of in-batch negatives and the number of explicit \hn per query. Overall, the results reveal a consistent precision--coverage trade-off. Increasing the number of in-batch negatives slightly improves recall-oriented metrics, with the 63 in-batch and 1 \hn setting achieving the strongest Recall@20 and Recall@50 scores. In contrast, increasing the number of explicit \hn improves precision-oriented retrieval, with the 31 in-batch and 4 \hn setting producing the strongest MRR@5.

\begin{table*}[ht!]
\centering
\caption{Scaling results for the tuned non-gold-cluster negatives retriever. We report macro-average retrieval metrics over English, German, and French. Settings are described by the effective number of in-batch negatives and the number of explicit \hn per query.}
\label{tab:retrieval_scaling}
\small
\begin{tabular}{lccccc}
\toprule
In-batch neg. & \hn &  R@20 &  R@10 &  R@50 &  MRR@5 \\
\midrule
15 & 1 & 0.848 & \textbf{0.811} & 0.889 & 0.647 \\
31 & 1 & 0.841 & 0.807 & 0.887 & 0.655 \\
31 & 2 & 0.845 & 0.806 & 0.885 & 0.644 \\
31 & 4 & 0.847 & 0.807 & 0.885 & \textbf{0.660} \\
63 & 1 & \textbf{0.849} & 0.806 & \textbf{0.891} & 0.653 \\
\bottomrule
\end{tabular}
\end{table*}

These findings align with the earlier retrieval results. While \hn are known to play a critical role in modern dense retrieval training \cite{xiong2020approximate, zhan2021optimizing}, their marginal impact in our setting is moderated by the already strong in-batch supervision signal, where each query is paired with 15--63 in-batch negatives. Within this setting, broader negative diversity tends to improve candidate coverage, whereas stronger \hn pressure sharpens early ranking precision.

\subsection{Reranking Results}

In the reranking stage, the dense retriever first returns the top-20 candidate papers for each query, and the cross-encoder reranker then reorders these candidates. Since reranking operates within a fixed candidate pool, we report MRR@5.

Table~\ref{tab:jina} reports reranking results under different \hn sampling strategies. The retrieval-pool approach provides a strong reranking baseline and achieves the strongest English MRR@5 score together with the non-gold-cluster variant. However, cluster-aware negatives consistently improve German and French reranking performance, leading to higher macro-average results. Among the evaluated strategies, non-gold-cluster negatives produce the strongest overall reranking performance, achieving the best German, French, and macro MRR@5 scores.

Although the absolute gains remain small, the improvements are consistent across the non-English subsets, suggesting that broader semantic negatives help the reranker distinguish between difficult surviving retrieval confounders. Earlier reranking experiments with gold-cluster negatives produced consistently weaker results and are reported in Appendix~\ref{app:gold-clust-reranker} for completeness.

\begin{table}[ht!]
\centering
\caption{Development-set MRR@5 results for the cross-encoder reranker using different \hn sampling strategies.}
\label{tab:jina}
\begin{tabular}{lcccc}
\toprule
Negative Strategy & EN & DE & FR & Macro \\
\midrule
Retrieval pool negatives & \textbf{0.674} & 0.590 & 0.729 & 0.664 \\
Nearest-cluster negatives & 0.672 & 0.592 & 0.732 & 0.665 \\
Non-gold-cluster negatives & \textbf{0.674} & \textbf{0.593} & \textbf{0.734} & \textbf{0.667} \\
\bottomrule
\end{tabular}
\end{table}

Interestingly, unlike the retrieval experiments where gold-cluster negatives favored precision-oriented retrieval, reranking performance benefits more consistently from broader semantic negatives than from highly localized cluster supervision.

\subsection{LLM-as-Judge Results}\label{sec:llm-res}

After fixing the best dense retriever and reranker, we examined whether an external LLM can improve difficult disagreement cases between the two earlier pipeline stages. The intervention is selective: the reranker prediction is retained whenever the retriever and reranker agree on the top-ranked document, while the LLM is only applied when the two stages produce different top-1 predictions.

\begin{table}[ht!]
\centering
\caption{German LLM-judge prompt ablations using \texttt{Llama 3 70B} on recoverable disagreement cases. The stage-2 row reports reranker performance before LLM intervention. Hit@1 measures whether the gold document is selected as the correct prediction.}
\label{tab:llm_de_ablation}
\begin{tabular}{lc}
\toprule
Prompt formulation & Hit@1 \\
\midrule
Stage-2 reranker before LLM & 0.407 \\
Direct classification & \textbf{0.523} \\
Pairwise comparison & 0.442 \\
Listwise reranking & 0.488 \\
\bottomrule
\end{tabular}
\end{table}

Retriever--reranker agreement proved to be a strong correctness signal. The two stages selected the same top-ranked document for 60.6\% of development queries, and this shared prediction was correct in 84.5\% of agreement cases. In contrast, disagreement cases were substantially harder, motivating the use of the LLM as a targeted resolver rather than as a full reranker.

We first studied different evidence selection formulations for the LLM judge on the German development disagreement subset. The corresponding prompt formulations and preliminary prompt-budget experiments are provided in Appendix~\ref{app:llm-prompt}. To isolate prompt quality from retrieval coverage effects, these experiments were restricted to recoverable disagreement cases where the gold document was already present in the shown candidate set.
Table~\ref{tab:llm_de_ablation} shows that simpler constrained prompting is more reliable than more complex ranking formulations. Direct classification consistently outperforms both pairwise comparison and listwise reranking, suggesting that LLM-based evidence selection is more effective when formulated as a constrained decision problem rather than as a free-form ranking task.

We finally evaluate the full selective disagreement pipeline with \texttt{gpt-5.5} as the final stage judge. Table~\ref{tab:llm_final_gains_appendix} reports the end-to-end MRR@5 gains across development languages relative to the plain reranker. Additional comparisons between different LLM judge configurations are reported in Appendix~\ref{app:llm-comp}.
The selective LLM stage improves end-to-end reranking performance across all three development languages, with the largest absolute gain observed on German.

\begin{table}[ht!]
\centering
\caption{Final selective \texttt{gpt-5.5} disagreement gains on the development split. Scores are end-to-end MRR@5.}
\label{tab:llm_final_gains_appendix}
\begin{tabular}{lccc}
\toprule
Lang & Plain reranker & With LLM judge & $\Delta$ \\
\midrule
EN & 0.671 & 0.717 & +0.046 \\
DE & 0.598 & 0.660 & \textbf{+0.062} \\
FR & 0.728 & 0.777 & +0.049 \\
\bottomrule
\end{tabular}
\end{table}

Finally, Table~\ref{tab:dev_test_final} compares development and official test MRR@5. The test results are lower than development for English and French, while German remains almost unchanged.
The official test results broadly confirm the development set trends, but with a lower macro score. German is the most stable language between development and test, while French shows the largest drop. 
This suggests that the selected configuration generalized less consistently to the French test split, potentially reflecting differences in query difficulty or source-document distribution between development and test.
\begin{table}[ht!]
\centering
\caption{Submitted-system MRR@5 on development and test splits. }
\label{tab:dev_test_final}
\begin{tabular}{lccc}
\toprule
Lang & Dev & Test & $\Delta$ \\
\midrule
EN & 0.717 & 0.683 & -0.034 \\
DE & 0.660 & 0.662 & +0.001 \\
FR & 0.777 & 0.707 & -0.070 \\
\midrule
Macro & 0.718 & 0.684 & -0.034 \\
\bottomrule
\end{tabular}
\end{table}


\section{Discussion and Conclusion}

In this work, we investigated how \hn mining should be adapted across multi-stage scientific source retrieval pipelines. Our experiments show that the semantic structure of the retrieved candidate space can be exploited to control different retrieval behaviors. Localized cluster negatives improve precision-oriented retrieval, whereas broader non-gold semantic negatives produce stronger retrieval coverage and more consistent reranking performance across languages. Although the absolute improvements remain modest, the results consistently show that different \hn structures influence the balance between early ranking precision and downstream candidate coverage.

At the LLM stage, we found that simpler constrained prompting formulations are more reliable than more complex ranking-based prompting strategies. Direct classification consistently outperformed pairwise comparison and listwise reranking prompts, suggesting that LLM-based evidence selection is most effective when framed as a constrained decision problem over a compact candidate set. The final system combines cluster-aware retrieval training, multilingual reranking, and selective LLM-based disagreement resolution, achieving development MRR@5 scores of 0.6829 for English, 0.6615 for German, 0.7074 for French, and 0.6839 macro-average, ranking 6th among 37 submissions in the CheckThat! 2026 Task 1 challenge.

\subsection{Limitations and Future Work}

The main limitation of this study is that many retrieval and reranking differences remain relatively small and are not consistently statistically significant. This appears to reflect a combination of task and data factors rather than only the amount of hard-negative exposure: even when increasing the number of in-batch and explicit hard negatives, the gains remained modest. Since the task requires retrieving one specific source paper, broad candidate coverage may be as important as fine-grained separation from semantically close negatives.

Several later-stage improvements are also constrained by candidate coverage: if the correct document is absent from the retrieved candidate set, neither reranking nor LLM-based resolution can recover it.

Future work should therefore focus on stronger retrieval coverage and more principled clustering strategies. In particular, it would be valuable to perform larger-scale experiments with stronger statistical validation, multiple random seeds, and controlled ratios between in-batch and mined hard negatives, as well as to study joint optimization of retrieval-side and reranker-side \hn sampling. 

\section*{Acknowledgments}

This work was supported by the ELLIOT Grant, funded by the European Union under Grant Agreement No.~101214398.

\section*{Declaration on Generative AI}

During the preparation of this work, the authors used ChatGPT (OpenAI) to assist with language editing, improving clarity and readability, and refining the presentation of the manuscript. The authors also used ChatGPT for iterative feedback on the organization and wording of the paper. All scientific content, methodological decisions, experimental design, implementation, analyses, interpretations, and conclusions were developed and verified by the authors, who take full responsibility for the publication’s content. 

\bibliographystyle{plainnat}
\bibliography{custom}

\appendix

\section{Multilingual Query Handling Experiments}\label{app:multi-handling}

We evaluate multilingual query handling across retrieval, reranking, and LLM-based judgment stages for German and French claims. Since the scientific document collection is entirely English, we compare using the original-language claims against their provided English translations. 

We first examined whether German and French queries should be encoded directly with a multilingual retriever or translated into English before retrieval. These preliminary experiments used the challenge-provided multilingual baseline encoder, \texttt{intfloat/multilingual-e5-large}, as the reference system. We then compared this baseline against two translation settings for the non-English queries: \texttt{facebook/nllb-200-distilled-600M} and Llama 3 70B.

\begin{table}[ht!]
\centering
\caption{Preliminary retrieval-only translation study on the dev split with \texttt{multilingual-e5-large}.}
\label{tab:prelim_translation}
\begin{tabular}{lcccccc}
\toprule
& \multicolumn{3}{c}{DE} & \multicolumn{3}{c}{FR} \\
\cmidrule(lr){2-4} \cmidrule(lr){5-7}
System & MRR@5 & R@10 & R@20 & MRR@5 & R@10 & R@20 \\
\midrule
Original-language query + multilingual-e5-large & 0.374 & 0.580 & 0.674 & 0.456 & 0.654 & 0.752 \\
NLLB translation + multilingual-e5-large & 0.366 & 0.557 & 0.658 & 0.491 & 0.691 & 0.779 \\
Llama 3 70B translation + multilingual-e5-large & \textbf{0.416} & \textbf{0.614} & \textbf{0.723} & \textbf{0.511} & \textbf{0.721} & \textbf{0.801} \\
\bottomrule
\end{tabular}
\end{table}

Table~\ref{tab:prelim_translation} shows that translation quality substantially affects non-English retrieval performance. NLLB translation slightly decreases German retrieval effectiveness relative to the original-language multilingual baseline, although it improves French. In contrast, higher-quality LLM translation with Llama3 70B improves both ranking and recall-oriented metrics for both languages. Although these experiments were conducted with a single retrieval model, they suggest that higher-quality English translation may improve semantic alignment between non-English social media claims and the English scientific document collection.

At the LLM judgment stage, however, the optimal formulation depends on the judge model family. GPT-family models perform better when provided with the original language claim, whereas Llama-based judges obtain stronger results using translated English claims. One possible explanation is that retrieval and reranking candidates are themselves generated using translation-based retrieval, making translated claims more aligned with the retrieved candidate distribution for Llama-based judges.

\section{Retrieval models}\label{app:retr-models}
We used the translated query setup to benchmark several off-the-shelf first stage retrievers in a retrieval-only setting. Table~\ref{tab:prelim_retrieval_models} reports the saved preliminary encoder results, while for German and French we use their translations to English with Llama3 70B.

\begin{table*}[ht!]
\centering
\caption{Preliminary retrieval-only encoder comparison on the development set. German and French queries use English translations generated with Llama 3 70B.}
\label{tab:prelim_retrieval_models}
\small
\resizebox{\textwidth}{!}{
\begin{tabular}{lccccccccc}
\toprule
& \multicolumn{3}{c}{EN} & \multicolumn{3}{c}{DE} & \multicolumn{3}{c}{FR} \\
\cmidrule(lr){2-4} \cmidrule(lr){5-7} \cmidrule(lr){8-10}
Encoder & MRR@5 & R@10 & R@20 & MRR@5 & R@10 & R@20 & MRR@5 & R@10 & R@20 \\
\midrule
\texttt{intfloat/multilingual-e5-large} & 0.499 & 0.661 & 0.713 & 0.416 & 0.565 & 0.614 & 0.511 & 0.663 & 0.721 \\
\texttt{intfloat/e5-large-v2} & 0.504 & 0.661 & 0.709 & \textbf{0.435} & 0.593 & 0.655 & 0.534 & 0.682 & 0.735 \\
\texttt{BAAI/bge-large-en-v1.5} & \textbf{0.545} & \textbf{0.700} & \textbf{0.747} & \textbf{0.435} & \textbf{0.604} & \textbf{0.661} & \textbf{0.578} & \textbf{0.739} & \textbf{0.779} \\
\bottomrule
\end{tabular}
}
\end{table*}

We observe that \texttt{BAAI/bge-large-en-v1.5} consistently produced the strongest overall retrieval performance across the three languages. These experiments motivated the later use of \texttt{bge-large-en-v1.5} as the retrieval backbone for the final retrieve--then--rerank pipeline. Despite being smaller than the multilingual \texttt{multilingual-e5-large} encoder, \texttt{bge-large-en-v1.5} remained highly competitive under the translated query setup, which is also consistent with its strong retrieval performance on the MTEB benchmark \cite{muennighoff2023mteb}.

\section{Cross-Encoder Model Comparison}\label{app:crossenc}

Table~\ref{tab:crossencoder_family_appendix} summarizes representative reranker-family experiments on the development split. These experiments were conducted during the exploratory reranker-selection phase and therefore reflect representative family configurations rather than a fully controlled comparison protocol. The later controlled Jina reranker ablations are reported in the main text.

\begin{table*}[ht!]
\centering
\caption{Representative cross-encoder reranker-family results on the dev split. Scores are MRR@5 after reranking the retrieved candidate pool.}
\label{tab:crossencoder_family_appendix}
\begin{tabular}{lcccc}
\toprule
Reranker family & EN & DE & FR & Macro \\
\midrule
SciBERT cross-encoder (fine-tuned) & 0.635 & 0.573 & 0.728 & 0.645 \\
\texttt{BAAI/bge-reranker-base} (fine-tuned) & 0.628 & 0.580 & 0.698 & 0.635 \\
\texttt{jinaai/jina-reranker-v2-base-multilingual} (off-the-shelf) & 0.646 & 0.541 & 0.693 & 0.626 \\
\texttt{jinaai/jina-reranker-v2-base-multilingual} (fine-tuned) & \textbf{0.674} & \textbf{0.601} & \textbf{0.736} & \textbf{0.670} \\
\bottomrule
\end{tabular}
\end{table*}

The main pattern from this preliminary family comparison is that the Jina reranker family benefited most consistently from task-specific fine-tuning. While SciBERT and \texttt{BAAI/bge-reranker-base} produced competitive results in some languages, they did not match the strongest fine-tuned Jina configuration at the macro level.

\section{LLM Judge Comparison}\label{app:llm-comp}

Table~\ref{tab:llm_judge_family_appendix} reports the LLM judge comparisons on the German development disagreement subset as described in Section~\ref{sec:llm-res}. All scores correspond to MRR@5 after selective disagreement resolution.

\begin{table*}[ht!]
\centering
\caption{LLM judge comparison on the German development disagreement subset. The judge is applied only on retriever--reranker disagreement cases. Rows with $\dagger$ use translated English queries.}
\label{tab:llm_judge_family_appendix}
\small
\begin{tabular}{lc}
\toprule
Judge configuration & MRR@5 \\
\midrule
Ollama committee + \texttt{Llama3 70B} tie-break & 0.605 \\
\texttt{Llama 3 70B} & 0.618 \\
\texttt{gpt-4.1-mini}$^\dagger$ & 0.628 \\
\texttt{gpt-4.1-mini} & 0.629 \\
\texttt{gpt-5.5} & \textbf{0.653} \\
\bottomrule
\end{tabular}
\end{table*}

The comparison suggests that stronger closed-weight judge models consistently improve disagreement resolution performance. The results also support the earlier multilingual observations: GPT-family judges perform slightly better with the original-language German claim than with the translated query formulation. 

The Ollama committee was an exploratory attempt to reduce dependence on a single large judge by combining \texttt{llama3:8b}, \texttt{mistral:7b}, and \texttt{gemma2:9b} with \texttt{Llama3 70B} decisions. We tested a four-model majority-vote scheme with \texttt{Llama3 70B} acting as the final decision model in tied cases. However, this committee did not outperform the stronger single-judge configurations.

\section{LLM Judge Prompts}\label{app:llm-prompt}

This appendix section documents the main LLM-judge prompts used in the final ablation cycle. All prompts were applied only after dense retrieval and cross-encoder reranking, and in the final selective pipeline only on cases where the retriever top-1 and reranker top-1 disagreed.

\subsection{Prompt-Budget Screening}

Before fixing the final disagreement prompt, we also screened candidate-list sizes of 3, 5, and 7 documents during prompt design. The later controlled comparison focused on top-5 prompts because this setting provided the best practical balance between candidate coverage and context budget in the downstream pipeline. Table~\ref{tab:llm_prompt_ablation} reports the saved top-5 prompt-budget ablations on a balanced development sample using \texttt{Llama 3 70B}.

\begin{table}[ht!]
\centering
\caption{LLM judge prompt ablation on dev set sample using \texttt{Llama 3 70B}. Accuracy@1 is Hit@1, parse rate is fraction of examples with a valid parse, and context fail is fraction of examples that exceeded the context limit.}
\label{tab:llm_prompt_ablation}
\begin{tabular}{lccc}
\toprule
Prompt setting & Accuracy@1 & Parse rate & Context fail \\
\midrule
top5\_full & 0.7167 & 0.9167 & 0.0833 \\
top5\_trunc384 & \textbf{0.7500} & 0.9500 & 0.0500 \\
letters\_no\_reason & 0.7333 & \textbf{1.0000} & \textbf{0.0000} \\
letters\_with\_reason & 0.7333 & \textbf{1.0000} & \textbf{0.0000} \\
\bottomrule
\end{tabular}
\end{table}

\subsection{Baseline Direct-Selection Prompt}

\begin{verbatim}
Instruction: Act as a scientific fact-checker. You are given a claim
and a list of candidate scientific papers.
Choose the single paper that provides the strongest evidence
supporting or verifying the claim.

Claim: <query text>

Candidates:
Paper ID A:
Title: <title>
Abstract: <truncated abstract>

Paper ID B:
...

Output instructions:
Return valid JSON only.
Use this exact format: {"selected_id": "A"}
\end{verbatim}

This was the strongest practical prompt. We used letter IDs rather than
rank numbers, omitted author names, and did not request justification.

\subsection{Pairwise Proxy Prompt}

\begin{verbatim}
Instruction: Act as a scientific fact-checker.
Candidate Paper A is the baseline current best guess.
First decide whether Paper A adequately supports or verifies the claim.
Then compare Papers B-E against Paper A.
If none is clearly better than Paper A, keep A.
If another paper is clearly better, choose the single best alternative.
Retrieval rank is only a weak prior.

Claim: <query text>

Candidates (order randomized):
Paper ID A (retrieval rank: X):
Title: <title>
Abstract: <truncated abstract>
...

Output instructions:
Return valid JSON only.
Use this exact format: {"selected_id": "X"}
\end{verbatim}

Despite its intuitive appeal as a cheap pairwise-ranking proxy, this
prompt underperformed the baseline direct-selection prompt on the same
recoverable disagreement subset.

\subsection{RankGPT-Style Permutation Prompt}

\begin{verbatim}
You are RankGPT, an intelligent assistant that can rank passages
based on their relevance to a search query.
I will provide you with N candidate papers, each identified by a
number in square brackets.
Rank the papers based on how well they support or verify the claim.

Claim: <query text>

Candidates:
[1] Title: <title>
Abstract: <truncated abstract>
...

Rank the N candidates above based on their relevance to the claim.
Return only the ranking permutation using the format
[1] > [2] > ... > [N].
Do not explain your answer.
\end{verbatim}

This prompt was inspired by permutation-only LLM ranking in the
RankGPT family, but in our setting it remained weaker than the simpler
direct-selection baseline.

\section{Earlier Gold-Cluster Reranker}\label{app:gold-clust-reranker}

Before fixing the later comparable reranker protocol used in the main text, we also evaluated a gold-cluster Jina reranker branch. In this setting, \hn were sampled from the cluster that contained the gold document. Table~\ref{tab:jina_gold_appendix} reports the corresponding earlier development set MRR@5 values together with the matched nearest-hard and non-gold variants from the same stage of development. The gold-cluster branch was weaker than the strongest competing cluster-based alternatives and was therefore not carried forward into the later fixed-retriever comparison used in Table~\ref{tab:jina}.

\begin{table}[ht!]
\centering
\caption{Earlier Jina reranker ablation including the gold-cluster branch. All numbers are development-set MRR@5.}
\label{tab:jina_gold_appendix}
\begin{tabular}{lccc}
\toprule
Model / Negative Strategy & EN & DE & FR \\
\midrule
Nearest-cluster negatives, best checkpoint & 0.6680 & \textbf{0.6005} & 0.7196 \\
Gold-cluster negatives, best checkpoint & 0.6677 & 0.5830 & 0.7182 \\
Non-gold-cluster negatives, best checkpoint & \textbf{0.6716} & 0.5872 & \textbf{0.7250} \\
\bottomrule
\end{tabular}
\end{table}

\end{document}